\documentclass[12pt]{article}

\pdfoutput=1
\usepackage{jheppub}
 \usepackage{etoolbox}% http://ctan.org/pkg/etoolbox
    \makeatletter
    \patchcmd{\maketitle}{\@fpheader}{}{}{}
    \makeatother
\usepackage{url}

\usepackage{xcolor}
\usepackage{fancyhdr}
\usepackage{amsthm}
\usepackage{float}
\usepackage{subfigure}
\usepackage{enumerate}
\usepackage{amsmath}
\newcommand{\be}{\begin{equation}}
\newcommand{\ee}{\end{equation}}
\newcommand{\bal}{\begin{aligned}}
\newcommand{\eal}{\end{aligned}}
\newcommand{\Log}{\mathrm{ln}}

 at10pt

\title{Entanglement Entropy in Causal Set Theory}

\author[a,b,c]{Rafael D. Sorkin,}
\author[a,c,d]{Yasaman K. Yazdi}

\affiliation[a]{Perimeter Institute for Theoretical Physics, 31 Caroline St. N., Waterloo ON, N2L 2Y5, Canada}
\affiliation[b]{Department of Physics, Syracuse University, Syracuse, NY 13244-1130, U.S.A.}
\affiliation[c]{Department of Physics and Astronomy, University of Waterloo, Waterloo ON, N2L 3G1, Canada}
\affiliation[d]{Department of Physics, University of Alberta, Edmonton AB, T6G 2E1, Canada}

\emailAdd{rsorkin@perimeterinstitute.ca, yyazdi@perimeterinstitute.ca}
 
\begin{document}

\abstract{Entanglement entropy is now widely accepted as having deep
  connections with quantum gravity.  It is therefore desirable to
  understand it in the context of causal sets, especially since they
  provide in a natural manner the UV cutoff needed to render
  entanglement entropy finite.
  Formulating a notion of entanglement entropy in a causal set is not
  straightforward because the type of canonical hypersurface-data on
  which its definition typically relies is not available.
  Instead, we appeal to the more global expression given in
  \cite{RDS1} which, for a gaussian scalar field, expresses the entropy
  of a spacetime region in terms of the field's correlation function
  within that region (its ``Wightman function'' $W(x,x')$).
  Carrying this formula over to the causal set, one obtains an entropy
  which is both finite and of a Lorentz invariant nature.
  We evaluate this global entropy-expression numerically for
  certain regions (primarily order-intervals or ``causal diamonds'')
  within causal sets of 1+1 dimensions.  For the causal-set counterpart
  of the entanglement entropy, we obtain, in the first instance, a
  result that follows a (spacetime) volume law instead of the expected
  (spatial) area law.
  We find, however, that one obtains an area law if one truncates
  the commutator function (``Pauli-Jordan operator") and
  the Wightman function by projecting out the eigenmodes
  of the Pauli-Jordan operator
  whose eigenvalues are too close to zero
  according to a geometrical criterion which we describe more fully
  below.
%%   Herein we evaluate this entropy for causal sets of 1+1 dimensions,
%%   specifically for order-intervals (``causal diamonds'') within the
%%   causal set.  We obtain in the first instance an entropy that obeys
%%   a (spacetime) volume law instead of the expected (spatial) area
%%   law.  We find, however, that one can obtain an area law by
%%   following a prescription for truncating the eigenvalues of a
%%   certain ``Pauli-Jordan'' operator and the projections of their
%%   eigenfunctions on the Wightman function that enters into the
%%   entropy formula.
  In connection with these
  results and the questions they raise,
  we also study the ``entropy of coarse-graining'' generated by
  thinning out the causal set, and we compare it with what one obtains
  by similarly thinning out a chain of harmonic oscillators, finding the
  same, ``universal'' behaviour in both cases.}
%\keywords{}

\pagestyle{plain}
%\begin{document}

\maketitle     % Just markar RETURN when error pops up!
% This elicits error, but don't delete it -- it is needed!

\thispagestyle{empty}
% \begin{document}

\section{Introduction}
Entanglement entropy is widely believed to be an important clue to a
better understanding of quantum gravity.  Beginning with the original proposal
that black hole entropy may be entanglement
entropy in whole or in part \cite{RDS}, and continuing through  the current surge of
interest excited by Van Raamsdonk's ideas on deriving the spacetime
metric
%% in AdS/CFT
from quantum entanglement \cite{vanram},
evidence has been accumulating that entanglement entropy has the
potential to unveil some of the mysteries surrounding the interplay
between the Lorentzian kinematics of general relativity and the
interference-laden dynamics of quantum theory.

Despite this history, it is only recently that a workable
definition of entanglement entropy has been formulated for causal sets
\cite{RDS1}.  While it turns out that a naive application of this
definition leads to a counter-intuitive spacetime-volume scaling (as
opposed to an entropy which scales as the spatial area in the limit of
small discreteness length), we will show below how to obtain the
anticipated area law by means of a suitable truncation scheme.  We will
also put forward an intuitive explanation of how the pre-truncation
volume-scaling arises, and of why it should
probably
be regarded as spurious from
the point of view of the continuum.
%% , and not a valid feature of causal sets.

\begin{figure}[h]
  \begin{center}
  \includegraphics[width=0.7\textwidth]{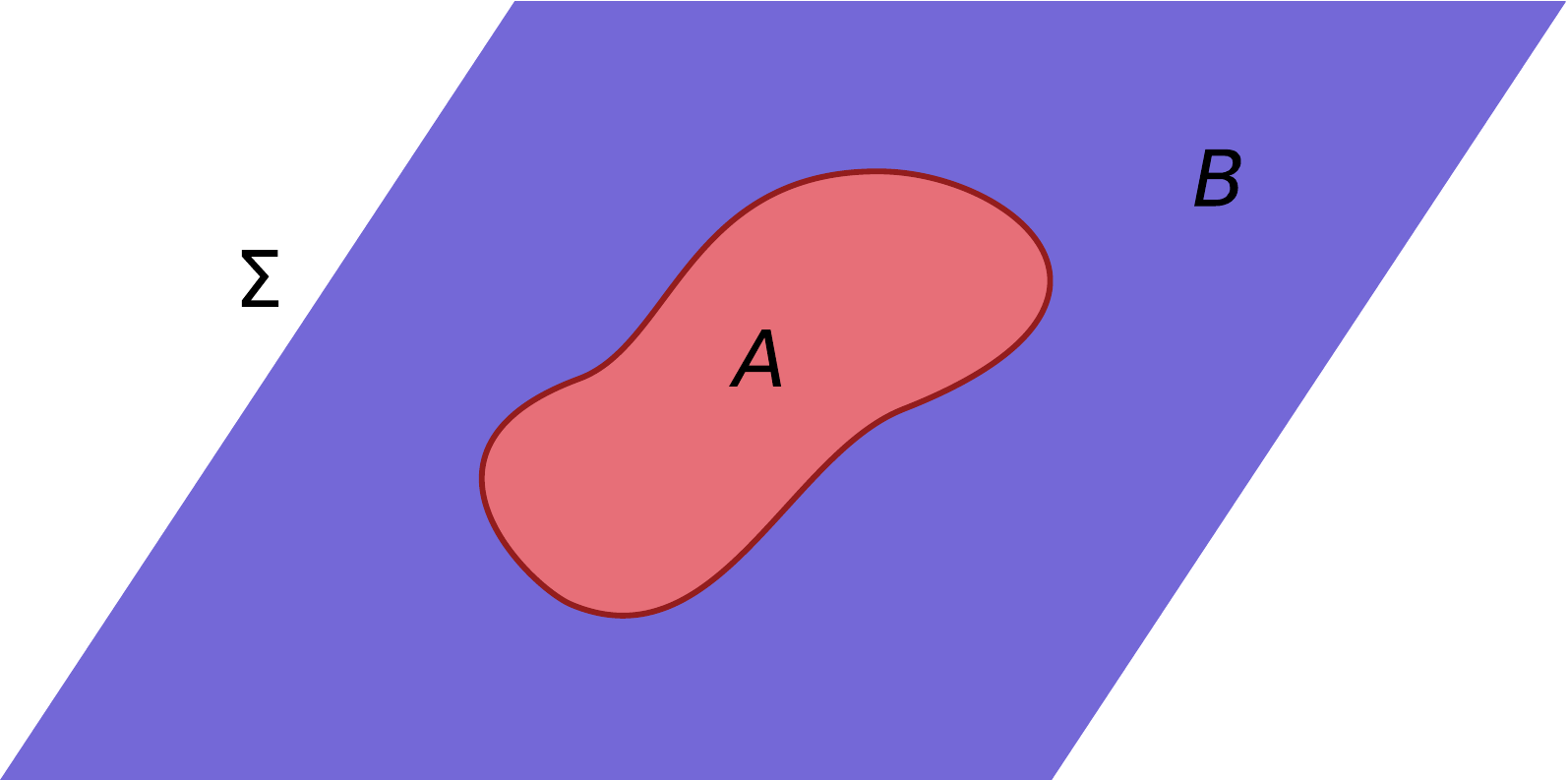}
  \caption{A hypersurface $\Sigma$ divided into two complementary subregions A and B.}
  \label{hypeAB}
  \end{center}
\end{figure}

Ordinarily, entropy is defined by the formula
\begin{equation}
  S = \text{Tr}\rho\ln\rho^{-1},
  \label{ee}
\end{equation}
where $\rho$ is a density matrix evaluated on a Cauchy hypersurface
$\Sigma$.  If $\Sigma$ is divided into two complementary subregions $A$ and $B$,
such as in Figure \ref{hypeAB},
then the reduced density matrix for subregion $A$ is
\begin{equation}
    \rho_A = \text{Tr}_B \rho.
    \label{red}
    \end{equation}
Substituting \eqref{red} back into \eqref{ee},
we get the entropy associated to region $A$ as
\begin{equation}
   S_A = -\text{Tr}\rho_A\ln\rho_A  \ ,  \label{e3}
\end{equation}
which can be designated as the entanglement
entropy between regions $A$ and  $B$
if the original density matrix $\rho$ was pure.
(We would of course get exactly the same answer if we instead traced over the
degrees of freedom of $A$ and computed $S_B$.)

This definition of entanglement entropy is unsatisfactory for more than
one reason.  First of all, it does not work for a causal set,
because we lack in that setting a notion of data on a hypersurface.
Even in the continuum moreover,
a hypersurface-based
definition of entanglement entropy seems questionable.  Essential to
getting a finite entanglement entropy is a UV cutoff, and a cutoff
referred to a spacelike surface has no reason to be covariant.  Two
partial Cauchy surfaces sharing the same boundary would then have no
reason to carry equal entanglement entropies, even if their domains of
dependence were the same.
It would
thus
be difficult to trust results
relying on a hypersurface-based cutoff,
such as area law scalings.
Also in a dynamical spacetime it would be
difficult to keep track of the entropy in a consistent way from one time
to another using such a cutoff.
%% , especially if the spacetime evolves non-trivially.
There is also the problem that strictly speaking (and especially in
interacting theories) quantum fields confined to a hypersurface do not
make sense, since they only become well defined when smeared in time.

Fortunately
%% Fortunately, however,
there exists a more
covariant definition of entanglement entropy which is formally
equivalent to (\ref{e3}) in a globally hyperbolic spacetime,
but which does not need to appeal to the notion of state on a hypersurface.
Moreover, this definition extends naturally to the
causal set context, allowing one to take advantange of
the Lorentz invariant cutoff which a causal set naturally provides.~\cite{RDS1}
So far, this definition has been developed for the theory of a
gaussian\footnote%
{In a gaussian theory Wick's rule holds, and the two-point function
 fully determines the theory.}
scalar field (also called a free scalar field in a quasi-free
state).
A definition for
a
fermionic field is soon to appear.
We review the definition for the scalar field
next.

\section{The Entropy of a Gaussian Field}
Let us review the definition of entanglement entropy in \cite{RDS1}. For
a more detailed review, we refer the reader to \cite{SSY}, and for the
full derivation to \cite{RDS1}.

The goal is to express $S$ directly in terms of the field correlators.
Consider first a single degree of freedom.
We introduce the Wightman and Pauli-Jordan matrices,
\be W=\left( \begin{array}{cc}
  \langle qq \rangle & \langle qp \rangle \\
  \langle pq \rangle & \langle pp \rangle  \\
  \end{array} \right)   \ee

and
 \be i\Delta=
  \left( \begin{array}{cc}
  0 & i  \\
  -i & 0  \\
  \end{array} \right) \ . \ee
The matrix $W$ corresponds in the field theory to
$W(x,x')=\langle 0|\phi(x)\phi(x')|0\rangle$,
while
$\Delta$ gives the imaginary part of $W$
and corresponds to the commutator function defined by
$i\Delta(x,x')=[\phi(x),\phi(x')]$.
Once we have these, we can express the entropy as a sum over the
solutions $\lambda$ of the generalized eigenvalue problem:
\be
    W \, v = i\lambda \; \Delta \, v  \ , \indent  \phantom{where} \indent \Delta \, v \neq 0  \ ,  \label{gee}
%%  W \, v = i\lambda \; \Delta \, v  \ , \indent     \text{where} \indent \Delta \, v \neq 0  \ .  \label{gee}
\ee
where the arguments of $W$ and $\Delta$ are restricted to the region in question.
In a causal set of finite cardinality,
or in the continuum with a mode cutoff,
%% Since we require a UV cutoff to render the entropy finite,
%% RDS: changed since there are ways to impose a cutoff without Delta
%% ceasing to be continuous. (as in our earlier "source of entropy" paper)
%
\eqref{gee}
becomes a matrix equation
where $W$ and $\Delta$ are finite dimensional matrices
and $v$ is a finite dimensional vector
in the image of $\Delta$.
%% (in some representation that imposes the cutoff, such as the causal
%% sets we consider in this paper)
The
specific
%% dimension of the
matrices and the
corresponding vector solutions $v$
will depend on
the nature of the cutoff
and
%% their detailed nature will depend
on the background spacetime or causal set on which they are defined.
The final expression for the entropy of the region is
\be
    S=\sum\limits_{\lambda} \lambda \, \ln |\lambda| \ . \label{s4}
\ee

In certain cases where we restrict $W$ and $\Delta$ to subregions within a
larger region or within an entire spacetime or causal set, \eqref{s4} can be
interpreted as an entanglement entropy.
In \cite{SSY}, \eqref{s4} was
applied to some examples in flat two-dimensional spacetimes,
and the conventional results for the scaling of the entropy with the
UV cutoff were found.
In the next section we will apply \eqref{s4} to the causal set counterparts of
the examples treated in \cite{SSY}.
In Section 4, we will apply it to obtain an entropy of coarse-graining
for a causal set.

\section{Causal Set Entanglement Entropy}
Causal set theory \cite{CS} is an approach to quantum gravity where the deep
structure of spacetime is discrete.  A causal set is a locally finite partially
ordered set.  Its elements are the ``spacetime atoms'', and its defining
order-relation is to be interpreted as the relation of causal or temporal
precedence between elements.  In the continuum the causal order and
the spacetime volume are enough to recover geometry.\footnote%
{In the discrete case, the volume of a region simply counts the number of
  elements in that region.}
An important feature of the theory is that in contrast to regular
lattices, causal sets are effectively Lorentz invariant.  For introductions to
causal set theory, we refer the reader to \cite{rev1, rev2}.

\begin{figure}[t]
\begin{center}
\includegraphics[width=0.7\textwidth]{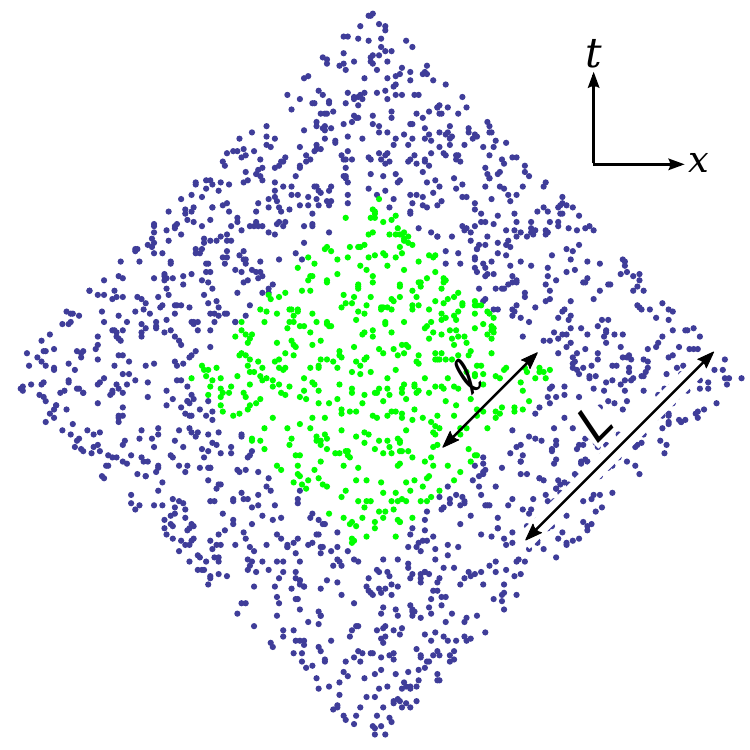}
\caption{Causal sets of two causal diamonds.}
\label{ee2d}
\end{center}
\end{figure}

Consider now a free gaussian scalar field living on a causal set which is well
approximated by a so-called causal diamond in a 2d Minkowski spacetime.  Using
the spacetime definition of entropy which was reviewed in the previous section,
let us compute the entropy associated to a smaller causal-set causal diamond
nested within the larger one.  Our setup is shown in Figure \ref{ee2d}, and the
entropy we will compute can be interpreted as that of the entanglement between
the small region and its ``causal complement''.  In less global terms, it is the
entanglement entropy between the ``equator'' of the smaller region, and its
complement within the Cauchy surface produced by extending this equator to the
larger region.

In the larger diamond, we use the $W$ of the Sorkin-Johnston vacuum
\cite{SJ1, SJ2, Yas}, $W_{SJ}$, which is the positive part of the operator $i\Delta$,
where
\be
     \Delta(x,y) = G_R(x, y) - G_R(y, x),
\ee
$G_R(x, y)$ being the retarded Green function.
For a (free) massless
scalar field, $G_R$ is simply related to the
causal matrix: $G_R=\frac{1}{2} C$, where C is the causal matrix,
\be
   C_{xy}:=
  \begin{cases} 1, & \text{if $x\prec y$}\\ 0, & \text{otherwise}
%%  \begin{cases} 1, & \text{if $n_x\prec n_y$}\\ 0, & \text{otherwise}
  \end{cases}
\ee

In solving \eqref{gee},
we restrict $W$ and $\Delta$ to elements within the
smaller diamond in Figure \ref{ee2d}, keeping
only the submatrices $W_{xy}$ and $\Delta_{xy}$ such that $x$ and $y$ are in the
smaller diamond.
In order to assess how the entropy scales with the UV cutoff,
we hold the ratio of the sizes of the diamonds fixed and vary
the number of elements sprinkled into them.
Then the UV cutoff
(given by the discreteness length-scale, which is in this case square root of the
density of elements)
is proportional to $\sqrt{N}$ where $N$ is the number of the causet
elements.
The UV cutoff is of course proportional to the square root of the number of
elements in both the larger and the smaller diamond;
we will use the number of
elements in the smaller diamond, $N_\ell$, to express it.

\begin{figure}[h]
  \begin{center}
  \includegraphics[width=0.9\textwidth]{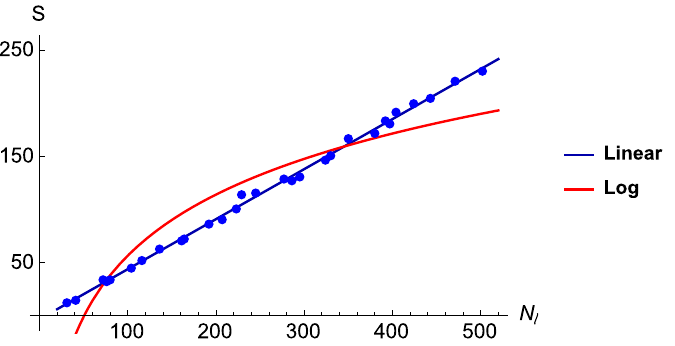}
  \caption{$S$ vs $N_\ell$ when $\ell/L=1/4$, along with best fits for linear and
    logarithmic functions. $N_\ell$ is the number of causet elements in the
    smaller diamond.}
  \label{sn02}

  \includegraphics[width=.9\textwidth]{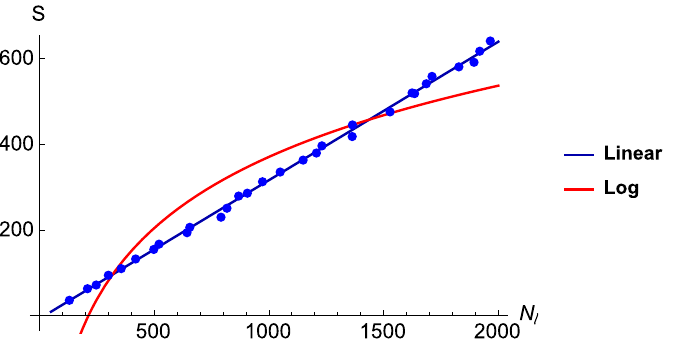}
  \caption{$S$ vs $N_\ell$ when $\ell/L=1/2$, along with best fits for linear and
    logarithmic functions. $N_\ell$ is the number of causet elements in the
    smaller diamond.}
  \label{sn04}
  \end{center}
\end{figure}

We find,
via numerical simulations,
that the entanglement entropy grows linearly with the number of elements
in the smaller diamond, thus obeying a spacetime-volume law\footnote%
{Notice that not only is this not an area law, but it is not even a spatial
  volume law. A spatial volume law would mean linear growth with $\sqrt{N}$,
  whereas the scaling that we obtain is linear in $N$.}!
The expectation, of course, was that (in $1+1$D) the entropy would scale
logarithmically with the UV cutoff
(which would mean logarithmic scaling with $\sqrt{N}$ and therefore with $N$
itself),
as in the continuum theory
\cite{SSY,acr}.
Furthermore, we find that the entropy in the causet is larger in
magnitude (values of order $100$) in comparison with the results in the continuum
(order 1 values).
Two examples of this linear scaling are shown in Figures
\ref{sn02} and \ref{sn04}, for $\ell/L=1/4$ and $\ell/L=1/2$, respectively. The
results fit $S=a N+b$ with $a= 0.46$ and $b=-3.20$ for $\ell/L=1/4$, and
$a=0.32$ and $b=-6.64$ for $\ell/L=1/2$.

%There is also a simple relation between the entanglement entropy and
%the ratio of the areas, $a/A$, of the diamonds in the causet. Figure
%\ref{aa} shows the results for a causal set with density
%$\rho=225$. The relation is quadratic and in this case fits
%$S=d\ (a/A)^2+b\ a/A+c$ with $\{d=-781.84, b=779.35, c=9.28\}$
%
%The relation between $S$ and $a/A$ can in general be expressed as
%\be
%\begin{split}
%S &\sim  b  N_{tot}  e (1- e)\\
%    &=  b  N_{1} ( N_{tot} - N_1)/N_{tot}\\
%    &=  b  N_1 N_2 /N_{tot}
%    \end{split}
%\ee
%
%where $N_1$ is the number of elements in the smaller diamond, $N_2$ the
%number of elements excluding $N_1$, and $N_{tot}$ the total number of
%elements. $b\sim 0.4$ and is independent of $N_{tot}$, so this entropy
%could also be thought of as some kind of entropy per pair. This result
%together with the linear scaling of entropy at fixed ratio $e$, lets us
%determine an expression for the slope of the linear
%scaling \footnote{Fay Dowker pointed this out.}: The slope is $\sim b
%(1- e)$.

% in the next sentence, we are listing 3 cases, not two: m>0, 4d, nonlocal G?

We also find that this spacetime-volume law persists
for the massive theory,
in $3+1$ dimensions,
and
when working with nonlocal Green functions such as that obtained from
the $G_R$ resulting from inverting the d'Alembertian defined in \cite{dal}.
See Appendix C of \cite{yasthesis} for more details on these cases.
This suggests that it is a generic feature of the direct
application of \eqref{gee}-\eqref{s4} to causal sets.

Whence comes this ``extra'' entropy?
The spectrum of $-i\Delta^{-1}W$ on a causal set necessarily has the same form
as in the continuum, in that its eigenvalues come in pairs of $\lambda$ and
$1-\lambda$.  However, many more of these pairs contribute to the entropy than
in the analogous continuum calculation.
A closer look at the spectrum of $i\Delta$
reveals how this happens.
In the definition \eqref{s4} it is
crucial that we exclude functions in the kernel of $i\Delta$, for which
$\lambda$ would not even be defined.
(Doing this also ensures that we have enough constraints to enforce the equations
of motion, so that only linearly independent degrees of freedom remain.)
While excluding the kernel is a simple
task for the continuum $i\Delta$, its meaning is not so straightforward for the
causal set $i\Delta$.
In the continuum, the number of ``zero-modes'' of $i\Delta$  is huge, but
in the causet it is much smaller.  Instead of strict zeroes one finds many small
but finite eigenvalues
that have no counterpart in the spectrum of the continuum $i\Delta$.
Even though these eigenvalues are very small, they can contribute a large amount
of entropy due to their being so numerous and to the inversion of $\Delta$ in
$-i\Delta^{-1}W$.

This observation leads to the idea that
(as suggested to us by Siavash Aslanbeigi)
these ``almost zero-modes'' of $i\Delta$ might be the source of the discrepancy\footnote
{We say ``discrepancy'' and not ``error'' since we don't wish to take a position
  on which, if either, of the two entropies is the ``correct'' one.}
between causet and continuum, and that they should be excluded from the entropy
calculation if one aims at agreement with the continuum.
If we start
removing
the contributions from
the smallest
eigenvalues $\tilde{\lambda}$ of $i\Delta$,
the scaling of the entropy with
the cutoff indeed becomes logarithmic.\footnote%
{We use $\tilde{\lambda}$ to refer to the spectrum of $i\Delta$, to avoid
  confusion with $\lambda$ which are the eigenvalues of $-i\Delta^{-1}W$ that go
  into \eqref{s4}.}
If the magnitude of the smallest eigenvalue
whose contribution
we keep is approximately
$\tilde{\lambda}_{min}\sim\sqrt{N}/4\pi$, then we get not only the expected
scaling-law but also the expected coefficient $1/3$ \cite{Cardy}.

\begin{figure}[h]
\begin{center}
\includegraphics[width=0.8\textwidth]{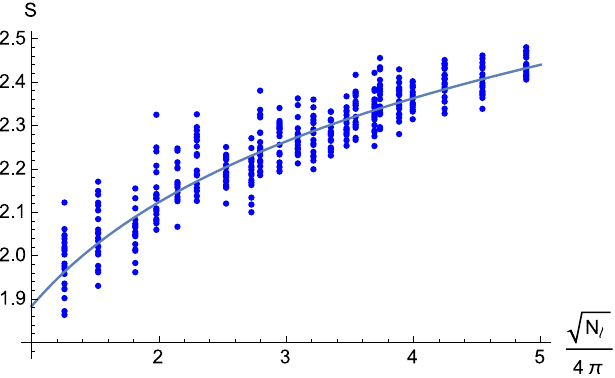}  % new from yasaman
\caption{S vs. $\sqrt{N_\ell}/4\pi$, after the spectrum of $i\Delta$ has been
  truncated such that $\tilde{\lambda}_{min}\sim\sqrt{N_L}/4\pi$ in the larger
  diamond and $\tilde{\lambda}_{min}\sim\sqrt{N_\ell}/4\pi$ in the smaller
  diamond.}
\label{trunc}
\end{center}
\end{figure}

An example of the logarithmic shape of the data points
after the truncation of $i\Delta$
is shown in Figure \ref{trunc}
for $\ell/L=1/2$.
In Figure
\ref{trunc}, the spectrum of $i\Delta$ has been truncated such that
$\tilde{\lambda}_{min}\sim\sqrt{N_L}/4\pi$ in the larger diamond and
$\tilde{\lambda}_{min}\sim\sqrt{N_\ell}/4\pi$ when the restriction is made to
the smaller diamond,
with contributions from the truncated eigenfunctions being projected out of $W$ as well.
(We first truncate both $\Delta$ and $W$ in the larger diamond ($\Delta$ being
the antisymmetric part of $W$). We then restrict both matrices to
the smaller diamond.  Call these restricted matrices $W^R$ and $\Delta^R$. We
then do a second truncation on them, based on the spectrum of $\Delta^R$.)
A fit
to $S=a\ln(x)+b$,
with $x$ being $\sqrt{N_\ell}/4\pi$ in the smaller diamond,
yielded $a=0.346 \pm 0.028$ and $b=1.883 \pm 0.035$,
consistent with the continuum value of $a=1/3$.
% $a=0.334$ and $b=1.901$.
It is worth emphasizing that the truncation has to be done
both in the larger diamond and in the smaller diamond.

%A fit to $S=b\ln(\sqrt{N_\ell})+c$ of this data was made and included in the
%figure and the best fit parameters were $b\sim0.329$ and $c\sim1.90$.\\

% this is figure 4

With hindsight we can understand why the magnitude of the smallest eigenvalue
has to be $\sim\sqrt{N}/4\pi$ for consistency with the continuum
results. The spectrum of $i\Delta$ in the continuum has dimensions of
area, while its spectrum in the causal set is dimensionless. This
dimensional observation, together with a comparison of the largest
eigenvalues of $i\Delta$  between continuum and causal set, shows that the
two spectra can be related by a density factor:
$\tilde{\lambda}^{cs}=\rho\tilde{\lambda}^{cont}$, where
$\rho=N_\ell/4\ell^2$.
Converting our $\tilde{\lambda}^{cs}_{min}$ to a
$\tilde{\lambda}^{cont}_{min}$
(in the small diamond),
we find
\be
\begin{split}
 \tilde{\lambda}^{cont}_{min}&=\tilde{\lambda}^{cs}_{min}/\rho\\
 &=\sqrt{N_\ell}/4\pi\rho\\
&\sim\frac{{\ell}^2}{\pi\sqrt{N_\ell}},
\end{split}
\ee
This is precisely\footnote%
{when we identify $\sqrt{N_\ell}$ as $n_{max}$.}
the  minimum eigenvalue which we retained in the continuum,
after imposing our cutoff on the wavelength of the eigenmodes of $i\Delta$.
This is reviewed in Appendix A.  Eigenvalues smaller than
$\tilde{\lambda}^{cs}_{min}$ thus correspond to solutions beyond the cutoff, and
are the ones we wish to exclude.

Another way to think of where the $\sqrt{N}/4\pi$ comes from is the following.
On one hand,
the causet provides a fundamental length
given (in 2d) by $\rho^{-1/2}$,
and in this sense
it serves as a ``low pass filter'' in relation to the continuum.
On the other hand,
in the continuum we know exactly
the relation between wavelength and eigenvalue
for eigenfunctions of $\Delta$ in a causal diamond.
If by means of this relation,
we convert a cutoff at wavelength $\rho^{-1/2}$
into a cutoff on the spectrum of $\Delta$,
we obtain the truncation rule stated above.
To the extent, then, that the asymptotic form of
the wavelength-eigenvalue relation is
%% asymptotically
universal
(as one might expect it to be),
one would expect to use
a qualitatively similar
%% the same
eigenvalue-cutoff,
not just for a causal diamond (order-interval),
but for a spacetime region of any shape.
%% RDS reinstated last sentence, with qualification

Truncating the spectrum of $i\Delta$ in the causal set by requiring its smallest
eigenvalue to be $\tilde{\lambda}_{min}\sim\sqrt{N}/4\pi$ reduces the size of
the spectrum from $\sim N$ to $\sim\sqrt{N}$.  Thus, a large number of these
approximate kernel-modes need to be eliminated if one wishes to recover an
area law.

\begin{figure}[h]
  \begin{center}
  \includegraphics[width=.99\textwidth]{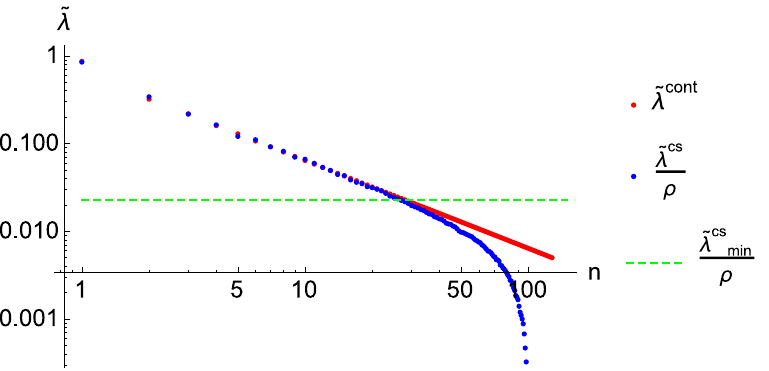}
  \caption{Comparison of the positive spectrum of $i\Delta$ in the continuum and causal set.
    The causal set has 200 elements and a density of 50.
    The green dashed line is where $\tilde{\lambda}^{cs}=\sqrt{N}/4\pi$.}
  \label{spec}
  \end{center}
\end{figure}

Figure \ref{spec} compares the positive spectrum of $i\Delta$ in the causal set
with that in the continuum, using a log-log plot.  The causal set comprises 200
elements sprinkled with a density of 50. The red dots are the continuum
eigenvalues, the blue dots those of the causet appropriately rescaled
by a factor of $1/\rho$
for the
comparison, and the green dashed line is at
$\tilde{\lambda}^{cs}=\sqrt{N}/4\pi$, where we would expect the causet
spectrum to end if it were to agree with the continuum. As one sees, the
eigenvalues above this line are in good agreement between causet and continuum,
but in very poor agreement below it.
In particular, there is a ``break''
in the causet spectrum
around where
the truncation has to be done.  Evidently,
this spectral feature could also be used as a guide for where to
truncate.
%% apply the truncation.

In summary,
we have seen
in these examples,
that one can recover
continuum-cum-cutoff behaviour from
the causal set by modifying
the condition $\Delta\,v\neq 0$ in \eqref{gee}
so as to exclude not only the strict zero-modes of $\Delta$,
but also its near-zero modes.
That is,
one identifies those modes $v$ for which
$i\Delta\,v=\tilde{\lambda}v$
with
$|\tilde{\lambda}|<\sqrt{N}/4\pi$
and projects them out
from W and $\Delta$
(doing so in both the bigger and smaller diamonds.)

%% In summary,
%% then, one can expect to recover
%% continuum-cum-cutoff behaviour from
%% the causal set by modifying the condition $i\Delta\, v\neq 0$ in \eqref{gee} to
%% $i\Delta\, v\neq \tilde{\lambda}_0 v$
%% when $|\tilde{\lambda}_0|<\sqrt{N}/4\pi$
%% and projecting out these near-zero parts from W as well.

\begin{figure}[h]
  \begin{center}
    \includegraphics[width=0.8\textwidth]{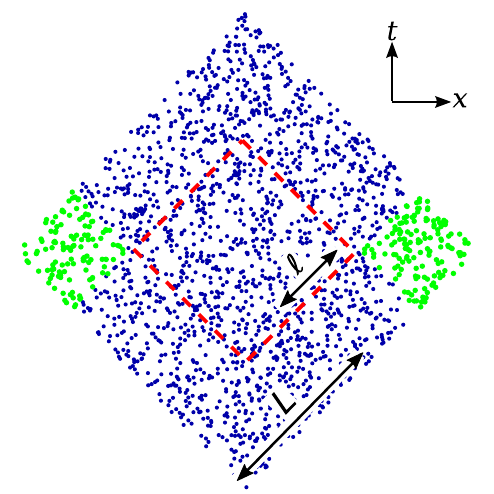}
    \caption{The domain of dependence of the complement of the ``Cauchy surface" in the causal diamond.}
%%    \caption{The domains of dependence of the complement of the ``Cauchy surface" in the causal diamond.}
  \label{cauch}
  \end{center}
\end{figure}

\begin{figure}[h]
\begin{center}
\includegraphics[width=0.8\textwidth]{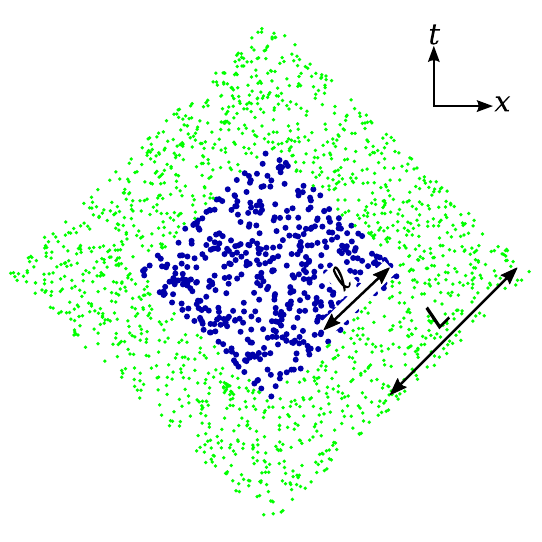}
\caption{The spacetime volume of the complement of the inner causal diamond.}
\label{vol}
\end{center}
\end{figure}

On the other hand, if one does not project out the near-zero modes, then
instead of the area-law familiar from the continuum, one encounters an
entropy that scales like spacetime volume.  Should this extra entropy be
regarded as physical, and if so how should it be interpreted?

In one way the resemblance with entanglement entropy is strong.  We
began in the larger diamond with a ``vacuum'' whose entropy vanished,
and we found on restricting it to a subregion (the smaller diamond) that
it induced there an ``impure state'' with nonzero entropy.  All this
resembles the entanglement-entropy associated with a bipartite system in
an overall pure state.  However the resemblance ends when one asks what
could play the role of the ``complementary subsystem'' to the smaller
diamond, in the sense of Figure \ref{hypeAB}.

Naively one might expect the complement of the inner diamond in
Figure \ref{ee2d} to be either the green subset of Figure \ref{cauch}
(the ``domain of dependence'' of the complement taken within the
``Cauchy surface'') or perhaps the green subset in Figure \ref{vol} (the
spacetime-complement of the smaller diamond itself).
But restriction of the scalar field to either of these subregions fails
(when one does {\it\/not\/} truncate the spectrum of $\Delta$)
to yield an entropy equal to that within the inner diamond, in conflict
with the fact that (bipartite) entanglement entropy can be computed
equally well from either one of the two complementary subsystems.

For the region of Figure \ref{vol} this shouldn't be surprising, because
the putatively complementary subregions are not mutually spacelike, and
the operators therein do not commute with one another.  Thus, even in
the continuum, one would not expect their entropies to be equal.
The situation is otherwise with the green subset of Figure \ref{cauch}.
In the continuum, it would be the ``causal complement'' of the smaller
diamond, and the two regions together would have the big diamond as
their domain of dependence or ``causal development''.  The
subregional entropies would be equal in that case.  That they are unequal
here is not an inconsistency because the operators from the two regions
taken together
don't
necessarily generate the full algebra of
operators of the big diamond.  In some sense, we are dealing
algebraically with a ``tripartite'' situation, but whether a meaningful
entanglement-entropy can be identified in this situation remains
unclear.
Note however, that if we perform the truncations in the causal set case,
then the entropy of the inner diamond indeed becomes equal to that of
its causal complement in Figure \ref{cauch}.

A final question arises in connection with the conjecture that black
hole entropy is, partly or wholly, entanglement entropy.  If this is
true then what role if any could an entropic volume-law play in relation
to area-law scaling of black hole entropy as usually understood?

\section{Entropy of Coarse-Graining}
In this section we study the entropy of coarse-graining by decimation and
blocking.  We study decimation in both causal sets and a
chain of harmonic oscillators, and blocking in a chain of harmonic
oscillators.  There is no known way to coarse-grain by blocking in a
causal set.  We use \eqref{s4} for the causal set calculation, and the formalism
of \cite{RDS2} for the oscillator calculations.

The Lagrangian for the chain of oscillators we consider is
\be
\mathcal{L}=\frac{1}{2}
 \left(\sum_{N=1}^{N_{max}} \hat{\dot{q}}_N^2-\sum_{N, M=1}^{N_{max}} V_{MN}\,\hat{q}_N \,\hat{q}_M\right)
 =\frac{1}{2}\sum_{N=1}^{N_{max}} [\hat{\dot{q}}_N^2-m^2 \hat{q}_N^2-k(\hat{q}_{N+1}-\hat{q}_N)^2],
\label{Lpho}
\ee
where k is the coupling strength between the oscillators, and in terms of the
spatial UV cutoff $a$, $k=1/a^2$ \cite{gold}. We set $k=10^6$. We consider the
massless theory with periodic boundary conditions and mass regulator\footnote%
{A mass regulator is introduced since the $m=0$ theory is IR divergent.
 See \cite{YKY} for more details.}
$m^2=10^{-6}$.

In coarse-graining by decimation, we iteratively remove $10\%$ of the
causet elements and oscillators. In the causet we remove each element
with probability $0.1$, and in the chain of oscillators we remove each
oscillator with probability $0.1$.  In more detail, at first we divide
the oscillators and causal set elements into two subsets: one subset
containing (approximately) $90\%$ of the oscillators and causet
elements, and the other (complementary) subset containing the remaining
(approximately) $10\%$. This division is done randomly, so the
oscillators in one subset may not necessarily have all of their nearest
neighbours from the full chain in that subset. Similarly, the elements
of the subset of the causal set are randomly chosen. Then we compute the
entanglement entropy between the two subsets. This is our first
(non-zero) entropy data point. Subsequently, we divide the subset
containing $\sim90\%$ of the original oscillators and causet elements
into two subsets containing $\sim90\%$ and $\sim10\%$ of them. We then
group this second $\sim10\%$ subset with the first $\sim10\%$ subset,
such that in terms of the original number of oscillators, our two
subsets at this second iteration contain $\sim81\%$ and $\sim19\%$ of
the total number of original oscillators and elements. The entanglement
entropy between these two subsets gives us our second (non-zero) entropy
data point. Similarly, each $n^{th}$ time we carry this out, we will
have $\sim0.9^n$ and $\sim1-0.9^n$ of the original number of oscillators
and causet elements in the two subsets whose entanglement entropy we
compute.

A simple
relation is obtained in both cases. The entropy depends quadratically on the
number of degrees of freedom (DoF's) remaining after coarse-graining. Initially,
when all DoF's are present, the entropy is 0. It rises and reaches a maximum
when about half of the DoF's remain, after which it drops, symmetrically, until
it reaches 0 again
when there are no more DoF's left.

\begin{figure}[H]
\begin{center}
\includegraphics[width=0.8\textwidth]{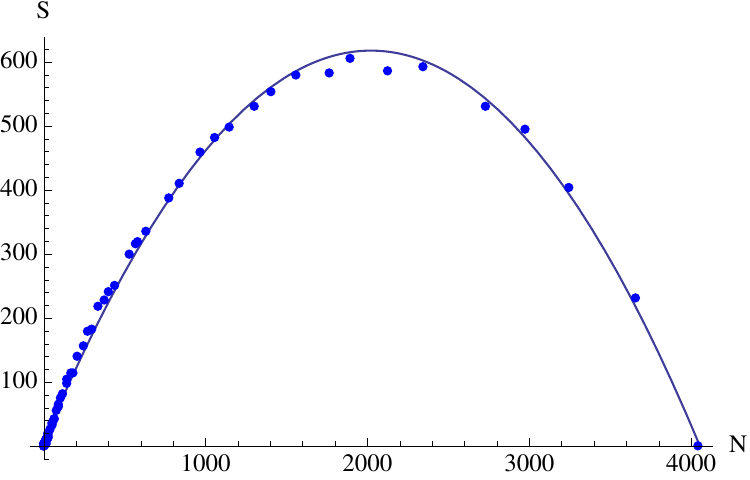}
\caption{$S$ vs. $N$ in a causet under coarse-graining (without truncating $i\Delta$ and $W$)
    by decimation: we remove elements with probability 0.1. }
\label{cscg}
\end{center}
\end{figure}

The causal set result without truncating $i\Delta$
and $W$
is shown in Figure \ref{cscg}, where the entropy is plotted versus the number of elements remaining
in the causal diamond. Initially the diamond contained 4048 sprinkled
elements
and had a density of $10.12$.
The results fit $S=a N^2+b N+c$ with $a=- 1.5\times 10^{-4},\, b=0.60,\, \text{and}\, c=7.0$.

\begin{figure}[]
\begin{center}
\includegraphics[width=0.85\textwidth]{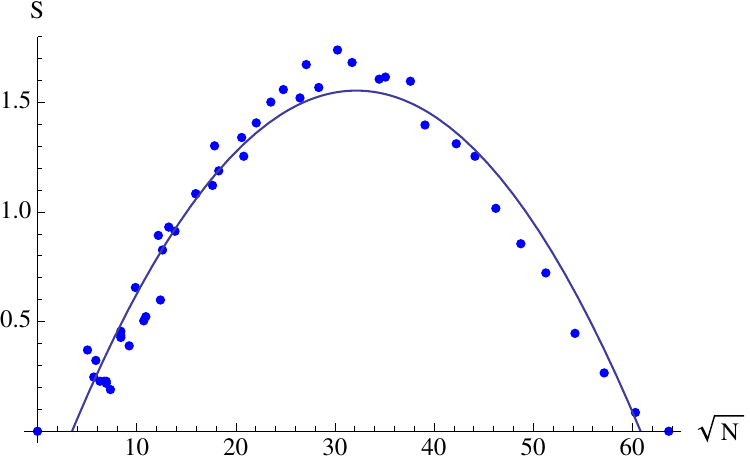}
\caption{$S$ vs. $\sqrt{N}$ in a causet under coarse-graining (with truncated $i\Delta$ and $W$) by decimation: we remove elements with probability 0.1. }
\label{cgtrunc}
\end{center}
\end{figure}

The causal set result with truncated\footnote%
{The first truncation in the full diamond is done identically to that
  used in Section 3. In other words, we make sure that $i\Delta$ does
  not have any eigenvalues with magnitude smaller than $\sqrt{N}/4\pi$,
  where $N$ is the total number of causal set elements. We similarly
  project out the contributions of the eigenfunctions corresponding to
  eigenvalues smaller than this value from $W$. The second truncation,
  however, is different from that used in Section 3. This is because our
  subset here is no longer a smaller diamond. Our subset in this case
  lives in the same larger diamond, so in our second truncation we use
  the same minimum eigenvalue of $\sqrt{N}/4\pi$ for the $i\Delta$
  restricted to the more dilute subset. $N$ is again the number of
  elements in the original full diamond (as opposed to the number of
  elements in the diluted subset). Similarly we project out their
  corresponding eigenfunctions from the restricted $W$ as well.}
$i\Delta$ and $W$
is shown in Figure \ref{cgtrunc},
where the entropy is plotted versus the square root of the number of elements
remaining in the causal diamond. Initially the diamond contained 4048 sprinkled
elements. The results fit $S=a N+b \sqrt{N}+c$ with $a=-0.0019,\, b=0.12,\, \text{and}\, c=-0.40$.

It should be noted that the DoF's in terms of which we get a parabolic
relation for the entropy of coarse-graining are different for the truncated and
full $i\Delta$ and $W$. For the full $i\Delta$ and $W$ the DoF's are counted by the number of
elements remaining in the diamond, $N$, and for the truncated $i\Delta$ and $W$ they are
counted by $\sqrt{N}$.

\begin{figure}[]
\begin{center}
\includegraphics[width=0.8\textwidth]{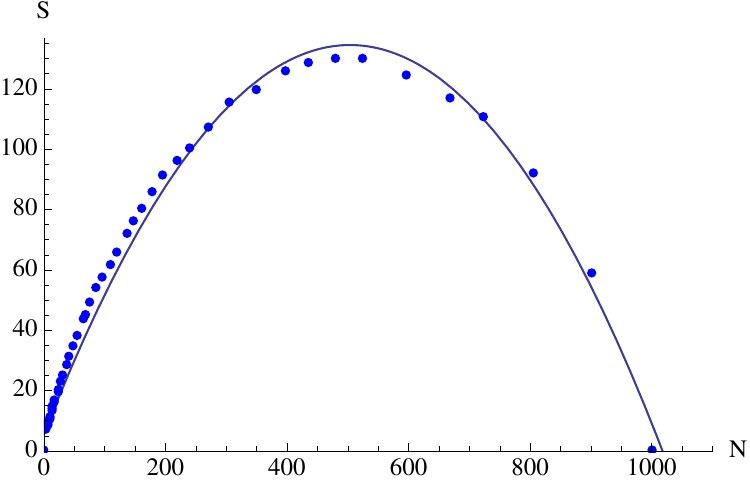}
\caption{$S$ vs. $N$ in a chain of oscillators under coarse-graining by decimation: we remove elements with probability 0.1.}
\label{hocg}
\end{center}
\end{figure}

The result for the chain of oscillators is shown in Figure \ref{hocg}, where the
entropy is plotted versus the number of oscillators remaining in the
chain. Initially the chain contained 1000 oscillators. The results fit $S=a N^2+b
N+c$ with $a=-5.1\times10^{-4},\, b=0.51, \,\text{and}\, c=5.4$.

\begin{figure}[]
\begin{center}
\includegraphics[width=0.8\textwidth]{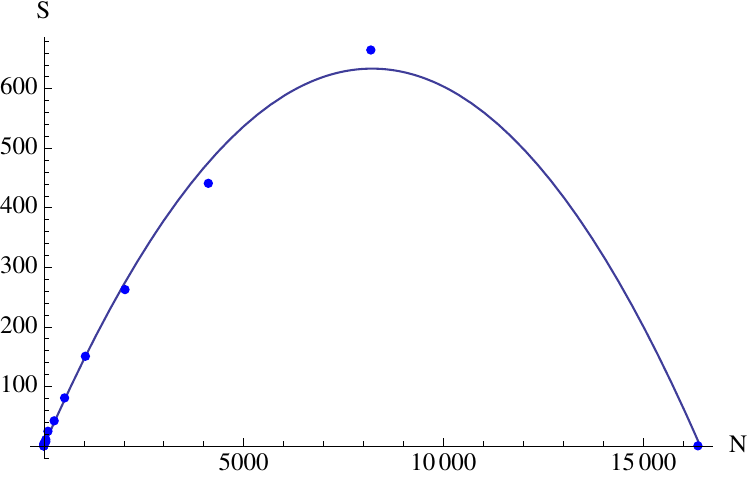}
\caption{$S$ vs. $N$ in a chain of oscillators under coarse-graining by blocking.}
\label{hocg2}
\end{center}
\end{figure}

In coarse-graining by blocking, we rewrite the $q_i$'s in terms of $Q_i^\pm$'s
defined as $Q_1^\pm\equiv(q_1\pm q_2)/2$, $Q_2^\pm\equiv(q_3\pm q_4)/2$, ... We
then discard all $Q^-$'s, thus reducing the DoF's by half. In the next iteration
we work in terms of $(Q^+_1\pm Q^+_2)/2$, $(Q^+_3\pm Q^+_4)/2$... and repeat.
The result for the entropy of coarse-graining by blocking in a chain of
oscillators is shown in Figure \ref{hocg2}. The entropy is shown versus the
number of oscillators remaining in the chain. Initially the chain contained
$2^{14}$ oscillators. The results fit $S=a N^2+b N+c$ with
$a=-9.4\times 10^{-6}, \, b=0.15,\, \text{and}\, c=-0.36$.

Thus entropy of coarse-graining by both decimation and blocking have led to a
parabolic dependence on the number of remaining DoF's, in our examples. Our
results suggest that this entropy of coarse-graining might have universal
properties that would be interesting to investigate further. We frequently deal
with coarse-grained versions of certain systems, and there seems to be an
entropy associated to this coarse-graining which has universal properties that
would be useful to understand.

Our choices of parameters ($\rho$ for the causal set, and $m$ and $k$
for the oscillators) in this section were arbitrary. As we change the
values of these parameters (as long as the UV cutoffs $\rho$ and $k$
remain large such that the asymptotic form of the entropy holds, and as
long as the mass, $m$, remains finite in order to avoid the infrared
divergence discussed in \cite{YKY}), the qualitative results of this
section do not change (as unpublished investigations have shown). The
magnitude of the maximum of the quadratic relation (Figures
\ref{cscg}--\ref{hocg2}) will, however, depend on  these
parameters. It would be interesting to analyze how this maximum scales
with each of the parameters. We defer this study to future work.

\section{Conclusions}

In the present paper,
we have studied (primarily by computer simulations) the entanglement entropy
of a free scalar field in causal sets well approximated by regions of
$1+1$D flat spacetime.  Initially we found unexpectedly that instead of
the conventional spatial area law (logarithmic scaling of entropy
with
%% the
UV cutoff), a spacetime-volume scaling was obtained.  We
attributed this difference between the causet and the continuum, to a
difference in the near-zero part of the spectrum of $i\Delta$.  With
this in mind, we identified, in the causet case, a minimum eigenvalue of
$i\Delta$ which answers to the fundamental discreteness scale embodied
in the causet itself.  And we found that when the spectrum of $i\Delta$
was truncated there
(and the contributions of these parts removed from $W$ as well),
the continuum area law was recovered.

With these findings, we are beginning to understand entanglement entropy
in causal set theory.  This is important for causal sets, of course, but
it also demonstrates an important point of principle, namely that the UV
cutoff needed to render entanglement entropy finite can be introduced
without doing violence to Lorentz symmetry. The way now seems open to
begin to address questions which hinge on understanding the entropy of
entanglement associated with black hole horizons, ultimately the
question whether most or all of the horizon entropy can be
traced to entanglement of one sort or another.

Work is also underway to find the entropy associated to the event
horizon of an ``observer'' in de Sitter spacetime \cite{yasds}. The retarded
Green function in a $3+1$D de Sitter causal set has recently been found
\cite{sumatidS} and makes possible the entropy calculation in that
setting. Also soon to appear is an application of the truncation scheme
presented in this paper to the Pauli-Jordan function derived from the
retarded Green functions which are inverses of the nonlocal causal set
d'Alembertians \cite{dionentropy}.

It is not yet known how the truncation procedure described in this paper
will generalize to higher dimensions,
and for arbitrary spacetime
regions. One speculation is that in $d$ spacetime dimensions, the
magnitude of the smallest eigenvalue is always of the order of
$N^{1/d}$. Another simple possibility for the generalization of the
truncation scheme is to include the largest $N^{(d-1)/d}$ eigenvalues
(possibly with a pre-factor that could depend on the shape of the
region) in $d$ spacetime dimensions. Both of these possibilities are
currently under investigation. Ultimately, however, to successfully
generalize this truncation scheme we would need to gain a better
understanding of the asymptotic (in the UV regime) nature of the
eigenfunctions and spectrum of $i\Delta$ in a general setting. Some
ideas in this direction (involving a conjecture that these eigenvalues
resemble a class of wavepackets and/or wavelets) are being pursued.

Whether the entropy-formula we have used will generalize to interacting
or non-gaussian theories, and if so how, is also an open question.  Any
such formula would necessarily have to supplement information from the
Wightman or ``two-point'' function with information corresponding to
correlators of higher degree.  Conversely the full complement of
$n$-point functions should determine the entropy uniquely, at least in
principle.  Perhaps a perturbative expansion about the gaussian case
could be formulated for interacting theories.
Or perhaps if the entropy-definition in the gaussian case could be
recast in path-integral form (for example, if it could be expressed as a
sum over the eigenvalues of a decoherence functional or other quantity
related to the path integral), the resulting expression could be
generalized to the non-gaussian case.

%% It is also not yet known how the definition of entropy we have used
%% could be generalized for interacting theories. Perhaps a perturbative
%% calculation along the same lines as the case for a gaussian state
%% (i.e. using Green functions) could be formulated for interacting
%% theories. Also, perhaps if the entropy definition could be recast in a
%% path integral form (for example, if the entropy could be expressed as
%% the sum over some function of the eigenvalues of a decoherence
%% functional or another quantity related to the path integral), it could
%% be generalized to the non-gaussian case.

The methods we have used in our simulations could also prove valuable in
a continuum context, as they illustrate how simulating entanglement
entropy via sprinkled causal sets can expedite calculations which would
otherwise be more tedious.

\section{Acknowledgements}
We thank Niayesh Afshordi, Achim Kempf, Sumati Surya, Mehdi Saravani, and Fay Dowker
for helpful discussions.
This research was supported in part by NSERC through grant RGPIN-418709-2012.
This research was supported in part by
Perimeter Institute for Theoretical Physics. Research at Perimeter
Institute is supported by the Government of Canada through the
Department of Innovation, Science and Economic Development Canada and by
the Province of Ontario through the Ministry of Research, Innovation and
Science.

\appendix
\section{Entanglement Entropy in Continuum Diamonds}
\subsection{Entanglement Entropy}
In this Appendix, we review the main results of \cite{SSY}.
\begin{figure}[h]
 \begin{center}
 \includegraphics[width=0.7\textwidth]{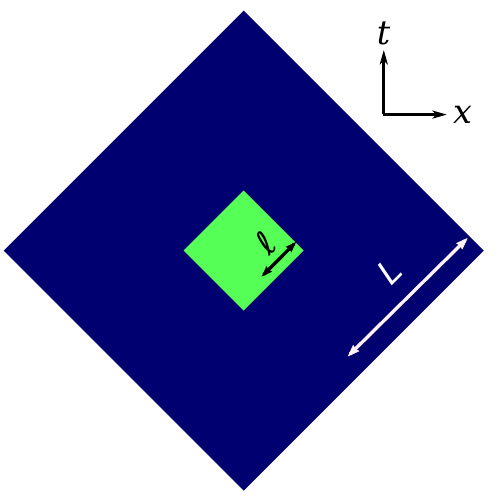}
 \caption{Two nested causal diamonds.}
 \label{2d}
 \end{center}
\end{figure}

We wish to use \eqref{s4} to compute the entanglement entropy of a
scalar field, resulting from restricting it to a smaller causal diamond
within a larger one in $1+1$D Minkowski spacetime. The setup is the
continuum analogue of Figure \ref{ee2d}, shown in Figure \ref{2d}. In
Minkowski lightcone coordinates
$u=\frac{t+x}{\sqrt{2}}$ and
$v=\frac{t-x}{\sqrt{2}}$,
\begin{equation}
   \Delta(u,v;u',v') = \frac{-1}{2} [ \theta (u-u')+\theta (v-v')-1], \
%%  i \Delta(u,v;u',v') = -\frac{ i}{2 } [ \theta (u-u')+\theta (v-v')-1] \ .
 \label{pjd}
\end{equation} and
 \begin{equation}
  W
   = -\frac{1}{4\pi} \Log|\Delta u \Delta v|
     - \frac{i}{4}\text{sgn}(\Delta u+\Delta v)\theta(\Delta u\Delta v)
     -\frac1{2\pi}\Log\frac{\pi}{4{L}}
     +\epsilon
     + \mathcal O\left(\frac{\delta}{{L}}\right),
\label{eq:SJtpcentrecomplete2}
\end{equation}

where $\epsilon\approx -0.063$ when $\ell\ll L$, and $\delta$
collectively denotes the coordinate differences $u-u',v-v',u-v',v-u'$. We set  $\frac{{\ell}}{L}=.01$.

We understand the properties of $\Delta$ and $W$ in this spacetime \cite{Yas}.

We represent $W$ and $\Delta$ as matrices in the eigenbasis of $i\Delta$ which consists of two sets of eigenfunctions:
\begin{align}
  f_k(u,v) &:= e^{-iku} - e^{-i k v}, & &\textrm{with } k = \frac{n \pi}{{\ell}}, \; n =\pm 1,\pm 2, \ldots\label{eq:SJfunctions3}\nonumber\\
  g_k(u,v) &:= e^{-iku} + e^{-i k v} - 2 \cos(k {\ell}), & &\textrm{with } k\in\mathcal{K},\end{align}
where
$\mathcal{K}=\left\{k\in\mathbb{R}\,|\,\tan(k{\ell})=2k{\ell} \ \textrm{and} \ k\neq0\right\}$.

The eigenvalues are ${\lambda}_k={\ell}/k$,
and the $L^2$-norms are $||{f}_{\small{k}}||^{2}=8{{\ell}}^{2}$
and $||{g}_{\small{k}}||^{2}=8{{\ell}}^{2}-16{{\ell}}^{2}{\cos}^{2}(k{\ell})$.

For the representation of  $W$, we computed $\langle f_k|W|f_{k'}\rangle$
and $\langle g_k|W|g_{k'}\rangle$. The terms $\langle f_k|W|g_{k'}\rangle$
vanish, making $W$ block diagonal in this basis.

The matrices representing $W$ and $\Delta$ are truncated to retain only
a finite number of eigenfunctions $f_k$ and $g_k$ up to a maximum value
$k = k_{max} = n_{max} \pi / \ell$.
Initial conditions described by functions with wavelengths
greater than $\lambda_{min}\sim 1/k_{max}$, can be expanded in terms of
these modes. A natural choice for the cutoff is therefore
$1/k_{max}$. In the calculations, we keep $\ell/L$ fixed and vary
$k_{max}$ (or equivalently $n_{max}$).
The eigenvalues are all of order one with absolute value
less than 3. All but a handful of the eigenvalue-pairs have values close to $1$ and $0$.
%Sample spectrum:
% \begin{figure}
%        \centering
%            \centering
%            \includegraphics[width=1.1\linewidth]{contspec}
%
%    \end{figure}

\begin{figure}
        \centering
            \centering
            \includegraphics[width=.75\linewidth]{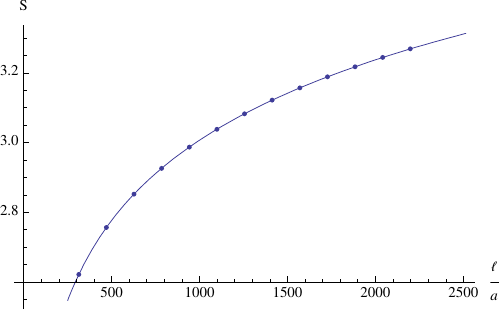}
            \caption{    Data points represent calculated values of $  S=\sum \lambda \, \ln |\lambda|$ in the continuum causal diamonds of Figure \ref{2d}.}
            \label{scont}
    \end{figure}

The obtained values of $S$, shown in Figure \ref{scont}, are fit almost perfectly
by the curve
\begin{equation}
      S = b \ln\left[ \frac{{\ell}}{a}\right] + c_1
\end{equation}
with
$b= 0.33277 $ and
$c_1=0.70782$.

\subsection{R\'{e}nyi Entropies}
We can extend the results of \cite{SSY} to include R\'{e}nyi entropies. The
spacetime definition of entropy given in \cite{RDS1} can be generalized
for R\'{e}nyi entropies of order $n$, $S^{(n)}$, in the following way:
\be
      S^{(n)}=\sum_\lambda\frac{-1}{1-n}\ln(\lambda^n-(\lambda-1)^n) \ ,
\label{Renyi}
\ee
% where $\lambda$ and $1-\lambda$ are solutions to the generalized
where $\lambda$ and $1-\lambda$ are solutions to the generalized eigenvalue problem \eqref{gee}. The spacetime we apply this formula to is
again Figure \ref{2d}. The expected result \cite{Cardy} is that the
entropies should scale as:
\be
   S^{(n)}=\frac{1}{6}(1+\frac{1}{n})\ln(\frac{\ell}{a})+c_n
\label{cftRenyi}
\ee
where $c_n$ are non-universal constants.

\begin{figure}[H]
  \begin{center}
  \includegraphics[width=0.8\textwidth]{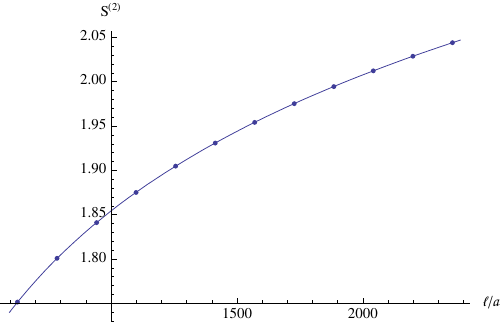}
  \caption{2nd order R\'{e}nyi entropy $S^{(2)}$ from \eqref{Renyi} vs. $\ell/a$ along with a fit to $S = b \ln\left[ \frac{{\ell}}{a}\right] + c$.}
  \label{r2}
  \end{center}
\end{figure}

\begin{figure}[H]
  \begin{center}
  \includegraphics[width=0.8\textwidth]{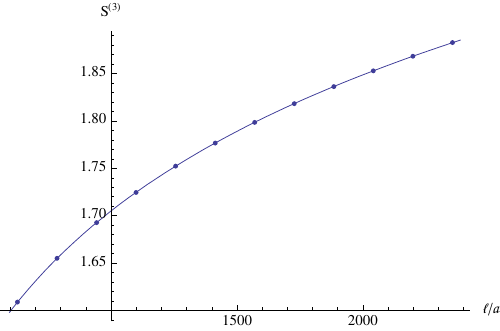}
  \caption{3rd order R\'{e}nyi entropy $S^{(3)}$ from \eqref{Renyi} vs. $\ell/a$ along with a fit to $S = b \ln\left[ \frac{{\ell}}{a}\right] + c$.}
  \label{r3}
  \end{center}
\end{figure}

Figures \ref{r2}-\ref{r3} show the results from \eqref{Renyi} for
$S^{(2)}$ and $S^{(3)}$.  There is good agreement between them
and \eqref{cftRenyi}, with more deviation present for
the higher order R\'{e}nyi entropies.
The scaling coefficients found from \eqref{Renyi} for $S^{(2)}$ to $S^{(10)}$  are:
\be
\{0.24961, 0.221498, 0.206892, 0.197726, 0.191411, 0.18682, 0.183354, 0.18066, 0.178517\},
\ee
to be  compared with those from \eqref{cftRenyi}:
\be
\{0.25, 0.222222, 0.208333, 0.2, 0.194444, 0.190476, 0.1875, 0.185185, 0.183333 \}.
\ee


\end{document}